\documentclass[showpacs,aps,twocolumn]{revtex4}
\usepackage{epsfig}
\usepackage{graphicx}
\usepackage{amsmath,amssymb,amsfonts}
\usepackage{array}
\usepackage{url}
\usepackage{multirow}
\usepackage{float}
\usepackage{lineno}
\usepackage{xspace}
\usepackage{ulem}
\usepackage{float}
\usepackage{comment}

\newcommand{\bef}{\begin{figure}}
\newcommand{\eef}{\end{figure}}
\newcommand{\bc}{\begin{center}}
\newcommand{\ec}{\end{center}}

\newcommand{\be}{\begin{equation}}
\newcommand{\ee}{\end{equation}}
\newcommand{\bea}{\begin{eqnarray}}
\newcommand{\eea}{\end{eqnarray}}

\def\ba{\begin{eqnarray}}
\def\ea{\end{eqnarray}}

\usepackage[usenames,dvipsnames]{color}
\definecolor{darkblue}{RGB}{0,0,196}
\usepackage[colorlinks=true,linkcolor=darkblue,citecolor=darkblue,urlcolor=darkblue]{hyperref}

\begin{document}
\title{Diffusion and fluctuations of open charmed hadrons in an interacting hadronic medium}

\author{Kangkan Goswami}
\author{Kshitish Kumar Pradhan}
\author{Dushmanta Sahu}
\author{Raghunath Sahoo\footnote{Corresponding Author: Raghunath.Sahoo@cern.ch}}
\affiliation{Department of Physics, Indian Institute of Technology Indore, Simrol, Indore 453552, India}

\begin{abstract}

Heavy quarks are excellent probes to understand the hot and dense medium formed in ultra-relativistic collisions. In a hadronic medium, studying the transport properties, e.g. the drag ($\gamma$), momentum diffusion ($B_{0}$), and spatial diffusion ($D_{s}$) coefficients of open charmed hadrons can provide useful information about the medium. Moreover, the fluctuations of charmed hadrons can help us to locate the onset of their deconfinement. In this work, we incorporate attractive and repulsive interactions in the well-established van der Waals hadron resonance gas model (VDWHRG) and study the diffusion and fluctuations of charmed hadrons. This study helps us understand the importance of interactions in the system, which affect both the diffusion and fluctuations of charmed hadrons.
\end{abstract}
\date{\today}
\maketitle
\section{Introduction}
    In a quest to explore the deconfined medium of partons and to create an early universe-like condition, ultra-relativistic heavy-ions have collided at the Relativistic Heavy Ion Collider (RHIC) and the Large Hadron Collider (LHC). Under such extreme conditions, an asymptotically free and locally thermalized system of deconfined partons is formed, called the quark-gluon plasma (QGP). Understanding the dynamics and interactions of such a medium is very interesting yet tricky, but estimating the thermodynamic and transport properties of the system formed in heavy-ion collisions can help us to understand the medium in a better way. One of the most effective probes to study the strongly interacting medium is the heavy quarks (HQs). This is because many heavier quark-antiquark pairs are produced in the initial hard scatterings. The relatively heavy quarks undergo Brownian motion in a medium of thermalized light-flavor quarks. Due to their large masses, their relaxation time is larger than the lifetime of the QGP, thus resulting in intermediate and high $p_T$ heavy quarks not getting thermalized in the medium \cite{Gossiaux:2009mk}. According to some phenomenological estimations, the QGP lifetime is estimated to be 4 - 5 fm/c at RHIC \cite{tHooft:2003lzk} and 10 - 12 fm/c at the LHC \cite{Foka:2016vta}. In contrast, the thermalization time of charm quarks is estimated to be of the order of 10-15 fm/c. Their mass is also much greater than the temperature of the system; thus, the probability of them being produced or annihilated in the medium is almost negligible. Hence, the HQs traverse the medium unaffected, but their momenta get modified due to their interaction with thermalized lighter quarks. These quarks hadronize around critical temperatures to form open-charm or open-bottom hadrons. However, it is crucial to notice that the momentum spectra of these hadrons undergo significant modification in the hadronic medium.
    Exploring the diffusion of these hadrons would help us to separate the contribution due to the hadronic sector from the deconfined phase. Thus, along with the charm quark, the study of drag and diffusion of $D^{0}$ meson ($c\bar u$) is of utmost importance. 

    The heavy-meson semi-leptonic decay produces electrons that serve as a mode to investigate the dynamics of the heavy meson. Generally, the nuclear suppression factor and elliptic flow of these electrons are analyzed to understand the drag and diffusion of the heavy mesons. Such experimental studies have been done at the RHIC and the LHC \cite{PHENIX:2006iih, Masciocchi:2011fu}. For the charm sector, the nuclear suppression factor and elliptic flow of $D^{0}$ mesons have been obtained at ALICE \cite{ALICE:2012ab, ConesadelValle:2012pfy}. Theoretically, many studies have also been done to explore the dynamics of heavy mesons in the hadronic medium. In a thermal bath of lighter hadrons, the $D^{0}$ mesons are significantly heavier, and their mass is much greater than the temperature of the system. One can exploit this fact to study the diffusion process of $D^{0}$ mesons in a dynamically changing medium, which can be mathematically described by the Fokker-Planck equation \cite{Svetitsky:1987gq}.

    In recent times, this idea has been utilized to study the diffusion of the open charmed state as well as the charm quark in thermalized hadronic and partonic media, respectively. In our previous work, we employed the Color string percolation model to study the drag and diffusion coefficients of the charm quark in the deconfined medium \cite{Goswami:2022szb}. HQ diffusion has also been studied extensively using perturbative QCD (pQCD) theory at leading order (LO) and next-to-leading order (NLO) \cite{Moore:2004tg, Svetitsky:1987gq, Caron-Huot:2008dyw}. There are also modifications to the pQCD theory, with addition of "hadronic" states in the deconfined phase, which shows a better agreement with other models \cite{vanHees:2004gq}. In ref.~\cite{Banerjee:2011ra, Brambilla:2020siz}, the authors have taken the lattice QCD (lQCD) approach to study the drag and diffusion of the charm quark. Moreover, works with T-matrix calculations explore the diffusion of charm quarks in the deconfined medium \cite{Riek:2010fk}. On the other hand, in the hadronic sector, the interaction of $D^{0}$ meson has also been studied using Born amplitudes \cite{Ghosh:2011bw}, and the spatial diffusion coefficient is found to have a smooth transition from the hadronic to the partonic medium. The drag and diffusion coefficients have also been evaluated in the framework of chiral perturbation theory (ChPT)~\cite{Das:2011vba} and heavy meson chiral perturbation theory \cite{Laine:2011is}. Moreover, the ideal hadron gas approach has been used to estimate the  $D^{0}$ meson diffusion in ref. \cite{Ozvenchuk:2014rpa}. In a recent work, the authors explore the magnetic field dependence on the $D^{+}$ meson diffusion using the fluctuation-dissipation theorem \cite{Satapathy:2022xdw}.

    Apart from studying the drag and diffusion of an open charmed hadron, one can also explore the melting of charmed hadrons in the medium to understand the medium in a better way. This may be explored by studying the fluctuations of open charmed hadrons. As the hot and dense medium expands violently, fluctuations in locally conserved quantities, e.g. net baryon number, electric charge, and strangeness show different behaviour in the hadronic medium as compared to a deconfined medium of quarks and gluons. In the hadronic medium, the baryon number carried by the particles is $\pm$ 1 or 0; however, for the QGP medium, it is only $\pm \frac{1}{3}$. Thus, a particle coming in or out of a sub-volume would produce quantitatively different fluctuations in the hadronic medium as compared to a QGP medium \cite{Jeon:2000wg}. These fluctuations show non-monotonic behavior at the phase boundary and hence can act as a potential probe to locate the phase boundary in the QCD phase diagram. Many have used this to explore the temperature at which there is a change in the degrees of freedom. Previously, studies \cite{Asakawa:2000wh, Jeon:2000wg} on fluctuations in the net baryon number and the electric charge have been done to explore the emergence of the deconfined medium and as a probe of chiral symmetry restoration. Similarly, net strangeness fluctuations and the appropriate ratios of their cumulants and cross-correlations give us an idea about the melting of strange mesons and baryons near the transition temperature~\cite{Koch:2005vg, Ejiri:2005wq, Bazavov:2013dta}. Fluctuations of net baryon number, electric charge, and strangeness have been explored broadly by the hybrid Polyakov–Nambu–Jona-Lasinio model \cite{Bhattacharyya:2017gwt, Bhattacharyya:2019qhm}, the Polyakov linear-$\sigma$ model \cite{Tawfik:2019kaz}, van der Waals hadron resonance gas model (VDWHRG) \cite{Vovchenko:2016rkn}, and the functional renormalization group approach \cite{Fu:2016tey}. Likewise, to understand the transition from charmed hadrons to charm quarks, one of the key methods is to investigate the melting point of the charmed hadrons. One can deploy the same strategy to understand the melting of open charm hadrons since it has been well established that charmonium states exist well above $T_c$ \cite{Bazavov:2014yba, Bellwied:2015lba}.
    However, the charm number fluctuations are rarely studied, making it a very intriguing topic of interest. In ref~\cite{Bazavov:2014yba}, Bazavov et al. have estimated the open charm fluctuations by using the lQCD theory. Thus, it would be interesting to see the results from other phenomenological models and their agreement with the lQCD results.

    The ideal hadron resonance gas (IHRG) model is a simple statistical model which successfully explains the lQCD results up to temperatures of 140-150 MeV. But near the transition temperature, the hadrons start to melt, and this model breaks down. There are various improvements to the IHRG model, such as the excluded volume hadron resonance gas (EVHRG) model, where the finite volume takes care of the repulsive interaction due to the hardcore radius of the hadrons. Recently, Vovchenko et al. \cite{Vovchenko:2016rkn} found that incorporating the van der Waals interaction between the hadrons improves the agreement with the lQCD results near the transition temperature. This van der Waals hadron resonance gas (VDWHRG) model has been used to explore various thermodynamic and transport properties of the hadronic matter \cite{Samanta:2017yhh, Sarkar:2018mbk, Pradhan:2022gbm, Behera:2022nfn, Poberezhnyuk:2019pxs, Pradhan:2023rvf}. In this work, we study the diffusion of $D^{0}$ meson, the net charm fluctuations, and their correlation with net baryon fluctuation, electric charge, and strangeness using the van der Waals hadron resonance gas model. In section \ref{vdw}, we briefly describe the formulation of the van der Waals HRG model. In section \ref{diff}, we present the results of diffusion of $D^{0}$ meson in an interacting hadronic medium. Finally, in section \ref{fluc}, we briefly discuss the melting of open charm hadrons and present our result. Finally, we discuss and summarize our results in section \ref{sum}.

\section{van der Waals hadron resonance gas model (VDWHRG)}
\label{vdw}
    The ideal HRG model is a thermally and chemically equilibrated statistical model consisting of non-interacting point-like hadrons. It can successfully reproduce results of various thermodynamic quantities from lQCD calculations \cite{HotQCD:2014kol}. In addition, the IHRG model can also be extended to a high baryochemical potential regime where the applicability of lQCD breaks down due to the fermion sign problem \cite{HotQCD:2014kol, Borsanyi:2013bia}. However, some disagreement with the lQCD data can be observed near the critical temperature. The major disagreements come while explaining the higher-order conserved charge fluctuations. A way out of this disagreement is to introduce interaction among hadrons at high temperatures. This is to take care of the qualitative features of the strong interaction that becomes much more significant as the temperature approaches $T_{c}$. To include the short-range repulsive interactions, one introduces a finite hardcore radius to all the hadrons, giving them a finite volume. This gives rise to the excluded volume HRG model (EVHRG). Although it improves the result near critical temperature, this model ignores the long-range attractive interactions.

    The van der Waals HRG model (VDWHRG) takes care of both attractive and repulsive interactions by introducing the $a$ and $b$ parameters, respectively.     In the VDWHRG model \cite{Vovchenko:2016rkn}, the authors have assumed that the interaction exists between baryon-baryon (antibaryon-antibaryon). However, the interactions between meson-meson, baryon-antibaryon, and meson-baryon (antibaryon) are not considered. One can safely exclude the short-range interactions between baryon-antibaryon as it is dominated by annihilation processes~\cite{Andronic:2012ut}. The interactions between mesons are neglected due to the fact that there is substantial suppression in the thermodynamic observables, which disagrees with lQCD results near the critical temperature at vanishing chemical potential. However, in recent years meson-meson repulsive interaction was included in the model by choosing a finite hardcore radius for the mesons, $r_{M}$~\cite{Sarkar:2018mbk}. Moreover, the attractive interaction among mesons leads to resonance formation, which is already present in the HRG model \cite{Venugopalan:1992hy, Savchuk:2019yxl} and hence not included in the formalism.

    Owing to the number fluctuation, the system created in a relativistic heavy-ion collision resembles the grand canonical ensemble (GCE). In the ideal HRG model, the grand canonical partition function of the $i^{th}$ hadronic species can be expressed as \cite{Andronic:2012ut}
    \begin{equation}
    \label{eq1}
    ln Z^{id}_i = \pm \frac{Vg_i}{2\pi^2} \int_{0}^{\infty} p^2 dp\ ln\{1\pm \exp[-(E_i-\mu_i)/T]\},
    \end{equation}
    where $g_i$, $E_i$, and $\mu_i$ are the degeneracy, energy and chemical potential of the $i^{th}$ hadron, respectively. The energy of the $i^{th}$ hadronic species is given as $E_i = \sqrt{p^2 + m_i^2}$, and $\mu_i$ can be further expanded in terms of the baryonic, strangeness, charge, charm chemical potentials and the corresponding conserved numbers as,
    \begin{equation}
    \label{eq2}
    \mu_i = B_i\mu_B + S_i\mu_S + Q_i\mu_Q + C_i\mu_C,
    \end{equation}
    
    where $B_{i}$, $S_{i}$, $Q_{i}$, and $C_{i}$ are, respectively, the baryon number, strangeness, electric charge, and charm quantum number of $i^{th}$ hadron. In the ideal HRG formalism, pressure $P^{id}_{i}$, and number density $n^{id}_{i}$  of an ideal hadron gas in the GCE can be written as,
    \begin{equation}
    \label{eq3}
    P^{id}_i(T,\mu_i) = \pm \frac{Tg_i}{2\pi^2} \int_{0}^{\infty} p^2 dp\ ln\{1\pm \exp[-(E_i-\mu_i)/T]\}
    \end{equation}
    \begin{equation}
    \label{eq5}
    n^{id}_i(T,\mu_i) = \frac{g_i}{2\pi^2} \int_{0}^{\infty} \frac{p^2 dp}{\exp[(E_i-\mu_i)/T]\pm1}
    \end{equation}
    
    To introduce van der Waals interaction, we start with the van der Waals equation of state in the canonical ensemble, which reads, 
    
    \begin{equation}
       \left( P + a\left( \frac{N}{V} \right)^2 \right)(V - bN) = NT,
       \label{eq7}
    \end{equation}
    
    where $P$, $N$, $V$, and $T$ are the pressure, number of particles, volume, and temperature of the system, respectively. The van der Waals parameters are $a$ and $b$, where $b$ is the eigen volume of the hadron given by
    \begin{equation*}
            b=\frac{16}{3}\pi r^{3}, 
    \end{equation*}
    $r$ is the hardcore radius of the hadron. In literature the parameters $a$ and $b$ are obtained either by reproducing the ground state of nuclear matter properties \cite{Vovchenko:2015vxa} or by fitting lQCD data of thermodynamic variables of hadronic gas \cite{Samanta:2017yhh, Sarkar:2018mbk}. In ref. \cite{Sarkar:2018mbk}, the parameters $a$ and $b$ are determined by simultaneously fitting the thermodynamic quantities obtained by lattice calculations, which, unlike earlier studies, also includes a finite radius for mesons making it a more realistic approach  to study the hadronic system. Hence, for our study, we choose the values as obtained in ref. \cite{Sarkar:2018mbk} as $a = 0.926 \rm$ $ \rm GeV. fm^{3}$ and for the parameter $b$, the hardcore radius for mesons and baryons (antibaryons) are taken as $r_{M} = 0.2$ fm and $r_{B(\bar{B})} = 0.62$ fm respectively. In the following formulation, $b$ is the repulsive parameter, where we take $b_{M}$ as the excluded volume for mesons and $b_{B(\bar{B})}$ as the excluded volume for baryons (antibaryons). Eq.~(\ref{eq7}) can be expressed in terms of number density, $N = n/V$ as, 
    \begin{equation}
        P(T,n) = \frac{nT}{1-bn} - an^{2}
    \end{equation}
    In the GCE, we can express pressure as~\cite{Vovchenko:2015vxa, Vovchenko:2015pya},

    \begin{equation}
        P(T,\mu) = P^{id}(T, \mu^{*}) - an^{2}
    \end{equation}
    where, $n$ is the number density calculated within the VDWHRG model and $\mu^{*}$ is the effective chemical potential given by,

    \begin{equation}
    \label{eq8}
    n(T,\mu) = \frac{\sum_{i}n_{i}^{id}(T,\mu^{*})}{1+b\sum_{i}n_{i}^{id}(T,\mu^{*})}
    \end{equation}

    \begin{equation}
    \label{eq11}
    \mu^{*} = \mu - bP(T,\mu) - abn^{2}(T,\mu) + 2an(T,\mu).
    \end{equation}
    
  The total pressure in the VDWHRG model can be written as,
    \begin{equation}
        P(T,\mu) = P_{M}(T,\mu) + P_{B}(T,\mu) + P_{\bar{B}},(T,\mu)
    \end{equation}
    where $P_{M}(T,\mu)$, $P_{B}(T,\mu)$, and $P_{\bar{B}}(T,\mu)$ are the pressure of the three subsystems defined in a VDW hadron gas, mesons with repulsive interaction, baryons and antibaryons with VDW interactions respectively. The pressure of these subsystems can further be expressed as,
    \begin{equation}
    \label{eq15}
    P_{M}(T,\mu) = \sum_{i\in M}P_{i}^{id}(T,\mu^{*}_{M})       
    \end{equation}
    \begin{equation}
    \label{eq16}
    P_{B}(T,\mu) = \sum_{i\in B}P_{i}^{id}(T,\mu^{*}_{B})-an^{2}_{B}(T,\mu)
    \end{equation}
    \begin{equation}
    \label{eq17}
    P_{\bar{B}}(T,\mu) = \sum_{i\in \bar{B}}P_{i}^{id}(T,\mu^{*}_{\bar{B}})-an^{2}_{\bar{B}}(T,\mu),
    \end{equation}

    where $M, B, \bar{B}$ stands for mesons, baryons and antibaryons respectively. $\mu^{*M}$ and $\mu^{*B(\bar{B})}$ are the effective chemical potential, for mesons and baryons(antibaryons) respectively. Considering vanishing chemical potential corresponding to electric charge, strangeness, and charm quantum number, i.e., $\mu_{Q}=\mu_{S}=\mu_{C}=0$, the modified chemical potential for mesons and baryon can be expressed as,

    \begin{equation}
        \mu^{*}_{M} = - b_{M} P_{M}(T,\mu)
    \end{equation}
        \begin{equation}
        \mu^{*}_{B(\bar{B})} = \mu_{B(\bar{B})}- b_{B(\bar{B})} P_{{B(\bar{B})}}(T,\mu) - ab_{B(\bar{B})}n^{2}_{B(\bar{B})} + 2an_{B(\bar{B})},
    \end{equation}
    where $n_{M}$ and $n_{B(\bar{B})}$ are the number density of mesons and baryons (antibaryons) in a VDW hadron gas and is given by Eq. (\ref{eq8}).
    
    Using the above VDWHRG formalism, we estimate the drag and diffusion coefficient of the  $D^{0}$ meson in an interacting hadron gas in the following section.

\section{Drag and Diffusion of $D^{0}$ meson}
\label{diff}
The charmed states are considerably heavier in a thermal bath of light hadrons, consisting mainly of pions, kaons, and protons. Hidden charm mesons like $J/\psi$ have much lower scattering cross-sections in the hadronic medium \cite{Mitra:2014ipa} as compared to that of open charmed mesons such as $D^{0}$ mesons. Thus, $D^{0}$ meson will diffuse sufficiently larger than $J/\psi$ in the hadronic medium. This affects the elliptic flow of $D^{0}$, while the $v_{2}$ of $J/\psi$ will remain unaffected, giving unfiltered information about the QGP phase. Thus, the interactions in the hadronic medium make the $D^{0}$ mesons an interesting probe to explore the hadron gas. Owing to the large mass difference, it has been established that one can reduce the Boltzmann transport equation or Boltzmann-Uehling-Uhlenbeck (BUU) equation to the Fokker-Planck equation to study the dynamics of the heavy meson in the hadronic medium. Although the Fokker-Planck and BUU methodologies exhibit significant differences, the drag and diffusion coefficient computed using these formalisms agree with each other considerably well \cite{Tolos:2016slr}.

The Fokker-Planck equation is given as,
\begin{equation}
\frac{\partial f(t,\textbf{p})}{\partial t} = \frac{\partial}{\partial p^{i}}\bigg[ (A^{i}(\textbf{p}) f(t,\textbf{p})) + \frac{\partial}{\partial p^{j}} ( B^{ij}(\textbf{p}) f(t,\textbf{p}))\bigg],
\end{equation}
where, $f(t,\textbf{p})$ is the momentum space distribution of $D^{0}$ meson with \textit{i,j}= 1,2,3 are the spatial indices. The collision kernels $A^{i}(\textbf{p})$ and $B^{ij}(\textbf{p})$ are given by \cite{Ozvenchuk:2014rpa},
\begin{equation}
A^{i}(\textbf{p}) = \int_{}^{}d\textbf{k}~ \omega(\textbf{p},\textbf{k})k^{i},
\end{equation}
\begin{equation}
B^{ij}(\textbf{p}) = \frac{1}{2} \int_{}^{}d\textbf{k}~ \omega(\textbf{p},\textbf{k})k^{i}k^{j}.
\end{equation}
Here, $\omega(\textbf{p},\textbf{k})$ is the collision rate of $D^{0}$ meson with initial momenta \textbf{p}, transferred momenta \textbf{k}, and final momenta \textbf{p}-\textbf{k}. Considering an isotropic medium and taking a static limit $ p \rightarrow 0$, where $p$ is the relative transverse momenta of the $D^{0}$ meson with the thermal bath. The collision kernels can be expressed in terms of drag and momentum diffusion coefficients as~\cite{Ozvenchuk:2014rpa},
\begin{equation}
    A_{i} = \gamma p_{i},
\end{equation}
\begin{equation}
    B_{ij} = B_{0}P_{ij}^{\bot}+B_{1}P_{ij}^{\parallel},
\end{equation}
where $\gamma$ is the drag coefficient, $B_{0}$ and $B_{1}$ are the transverse and longitudinal momentum diffusion coefficients respectively. $P_{ij}^{\bot}$ and $P_{ij}^{\parallel}$ are the perpendicular and parallel components of the projection operator. The $D^{0}$ meson undergoes Brownian motion in a thermal bath of lighter hadrons losing its momenta. The average momenta of the $D^{0}$ meson in the hadronic medium can be expressed as\cite{Torres-Rincon:2013nfa}, 
\begin{equation}
\label{eq17}
\langle p \rangle = \frac{ \int_{-\infty}^{\infty} dp ~ p ~f(t,p)}{\int_{-\infty}^{\infty} dp~f(t,p)} = p_{0}~e^{-\frac{t}{\tau}}
\end{equation}
where, $\tau$ is the relaxation time of the $D^{0}$ meson and $p_{0}$ is the initial momenta. The relaxation time of the $D^{0}$ meson is related to the drag coefficient as $\tau = 1/\gamma$ \cite{Torres-Rincon:2013nfa}.

\begin{figure}[H]
    \centering
    \includegraphics[width=1\linewidth]{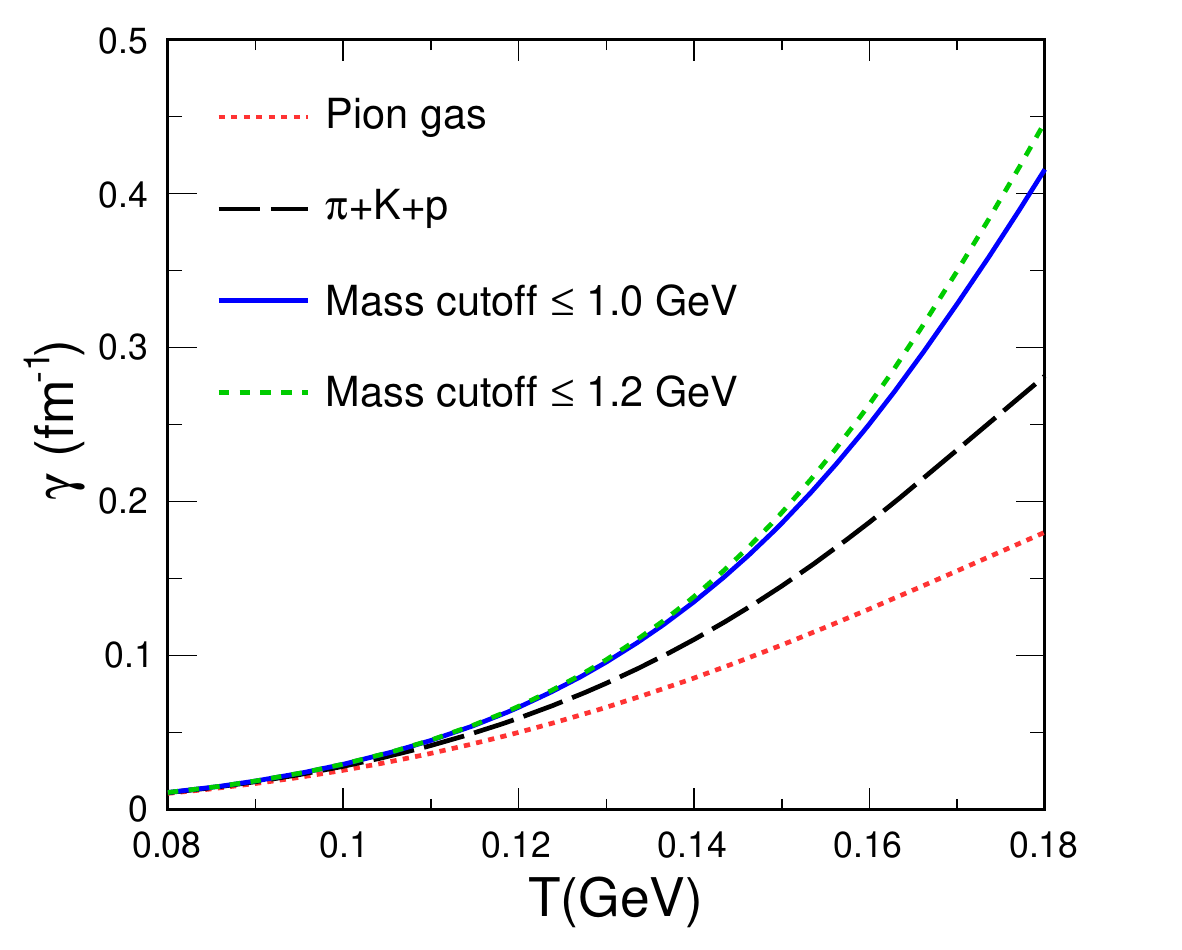}
    \caption{Drag coefficient of $D^{0}$ meson as a function of temperature at $\mu_{B}=0~\rm{GeV}$ with different mass cutoff.}
    \label{fig1}
\end{figure}

In accordance with the widely used relaxation time approximation \cite{Roberts:2000aa}, 
the relaxation time for $D^{0}$ meson in a hadron gas can be expressed as,
\begin{equation}
\label{eqtau}
    \tau^{-1} = \sum_{j} n_{j} \langle \sigma_{j}v_{j} \rangle,
\end{equation}
where $n_{j}$ is the number density of $j^{th}$ hadronic species. $\sigma_{j}$ and $v_{j}$ are the cross-section and relative velocities between $j^{th}$ hadronic species and $D^{0}$ meson. Their thermal average can be approximated as \cite{Kadam:2015xsa},

\begin{multline}
    \langle \sigma_{j} v_{j} \rangle = \frac{\sigma_{Dj}}{8Tm_{D}^{2}m_{j}^{2}K_{2}(\frac{m_{D}}{T})K_{2}(\frac{m_{j}}{T})} \int_{(m_{D} + m_{j})^{2}}^{\infty} \\ ds\frac{s-(m_{D}-m{j})^{2}}{\sqrt{s}} (s-(m_{D}+m_{j})^2)K_{1}(\frac{\sqrt{s}}{T}). 
\end{multline}
Here, $m_{j}$ is the mass of $j^{th}$ species of hadron and $m_D$ is the mass of $D^0$ meson. $s = (p_{D} + p_{j})^2$ is the Mandelstam variable, and $K_{n}$ is the modified Bessel function of $n$th order. Following the Ref. \cite{Ozvenchuk:2014rpa, He:2011yi}, $Dm \rightarrow Dm$ and $DB(\overline{B}) \rightarrow DB(\overline{B})$ elastic scattering cross section is taken as $\sigma = 10$ mb and $\sigma = 15$ mb respectively, where $m$, $B$, and $\overline{B}$ are mesons, baryons, and antibaryons respectively.   On estimating $\tau^{-1}$, we obtain the drag coefficient, $\gamma$ as a function of temperature. In an isotropic medium and the static limit ($p\rightarrow0$), the transverse and longitudinal momentum diffusion coefficient follows, $B_{0}$ = $B_{1}$. It describes the broadening of momentum spectra of final state hadrons. From Einstein's relation, we can express momentum diffusion coefficients in terms of drag coefficient, temperature, and mass of $D^{0}$ meson as \cite{Tolos:2016slr},
\begin{equation}
    B_{0} = \gamma m_{D}T,
\end{equation}

Finally, we estimate the spatial diffusion coefficient, $D_{s}$, to understand $D^{0}$ meson diffusion in coordinate space. The mean quadratic displacement of $D^{0}$ meson as a function of time is given as \cite{Torres-Rincon:2013nfa},
\begin{equation}
    \langle (x(t) - x(t=0))^{2} \rangle = 2D_{s}t
\end{equation}
It can be understood as the speed of $D^{0}$ diffusion in space in a hadronic medium. Under the static limit, $D_{s}$ can be obtained as,
\begin{equation}
    D_{s} = \frac{T}{m_{D} \gamma}.
\end{equation}

\begin{figure*}
    \centering
    \includegraphics[width=0.45\linewidth]{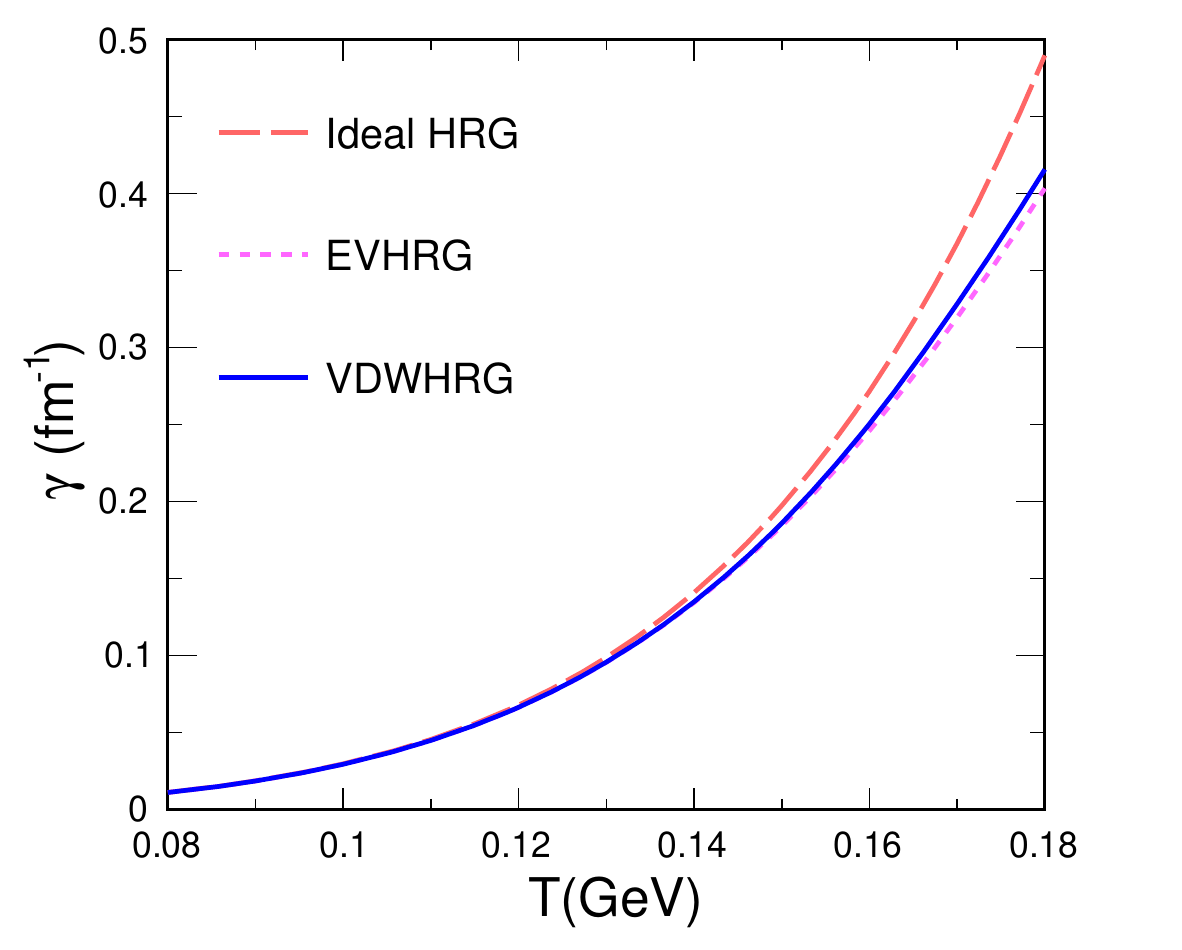}
    \includegraphics[width=0.45\linewidth]{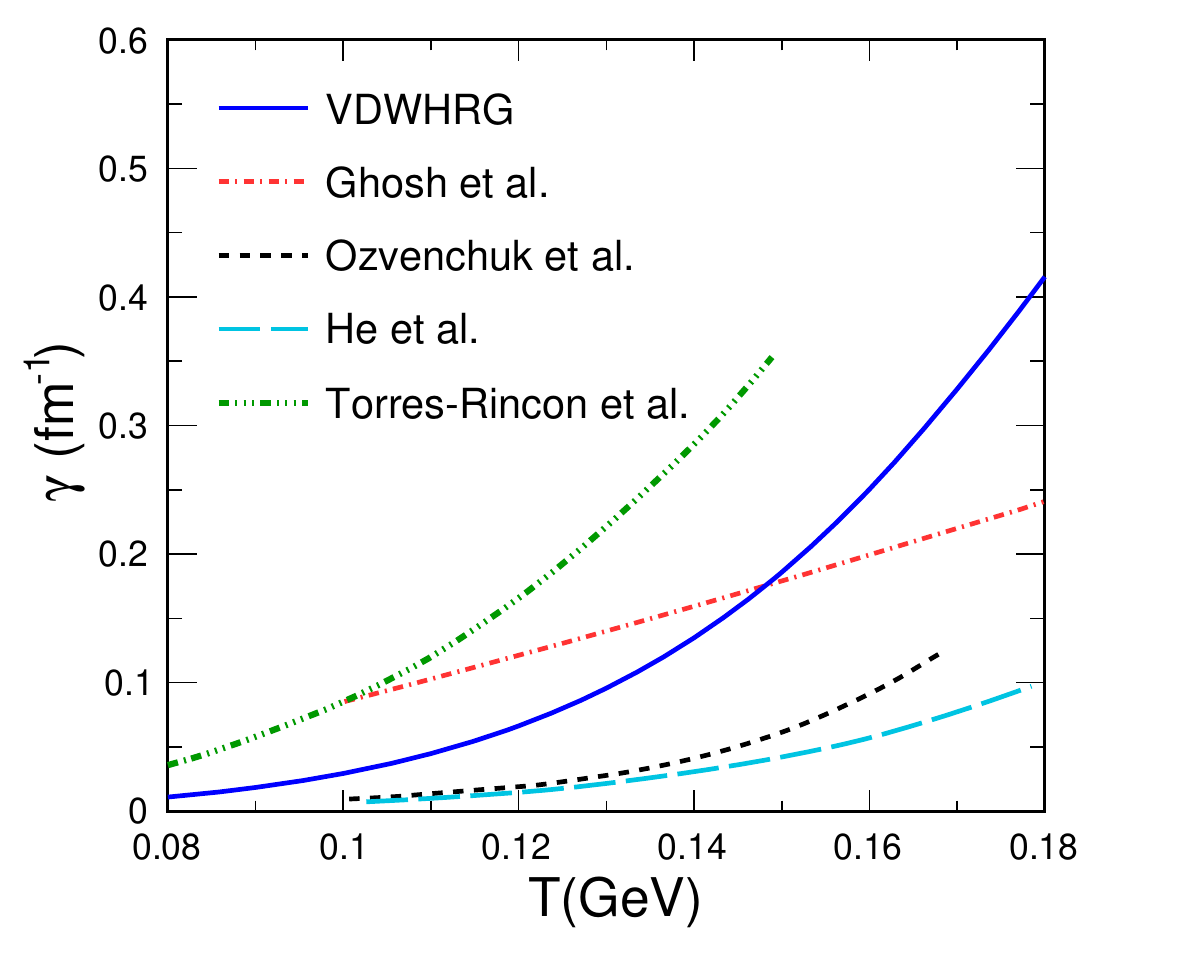}
        \caption{Variation of drag coefficient of $D^{0}$ meson with temperature at $\mu_{B}=0~\rm{GeV}$. A comparison among the Ideal HRG, EVHRG, and VDWHRG models (left). A comparison between our result and different phenomenological models (right). The green dashed-dotted-dotted line is taken from Ref.~\cite{Torres-Rincon:2021yga}, the red dot-dash line is obtained from Ref.~\cite{Ghosh:2011bw}, the black dotted line is the result from Ref.~\cite{Ozvenchuk:2014rpa}, and the cyan dashed line is from Ref.~\cite{He:2011yi}. }
    \label{fig2}
\end{figure*}

\begin{figure*}
    \centering
    \includegraphics[width=0.45\linewidth]{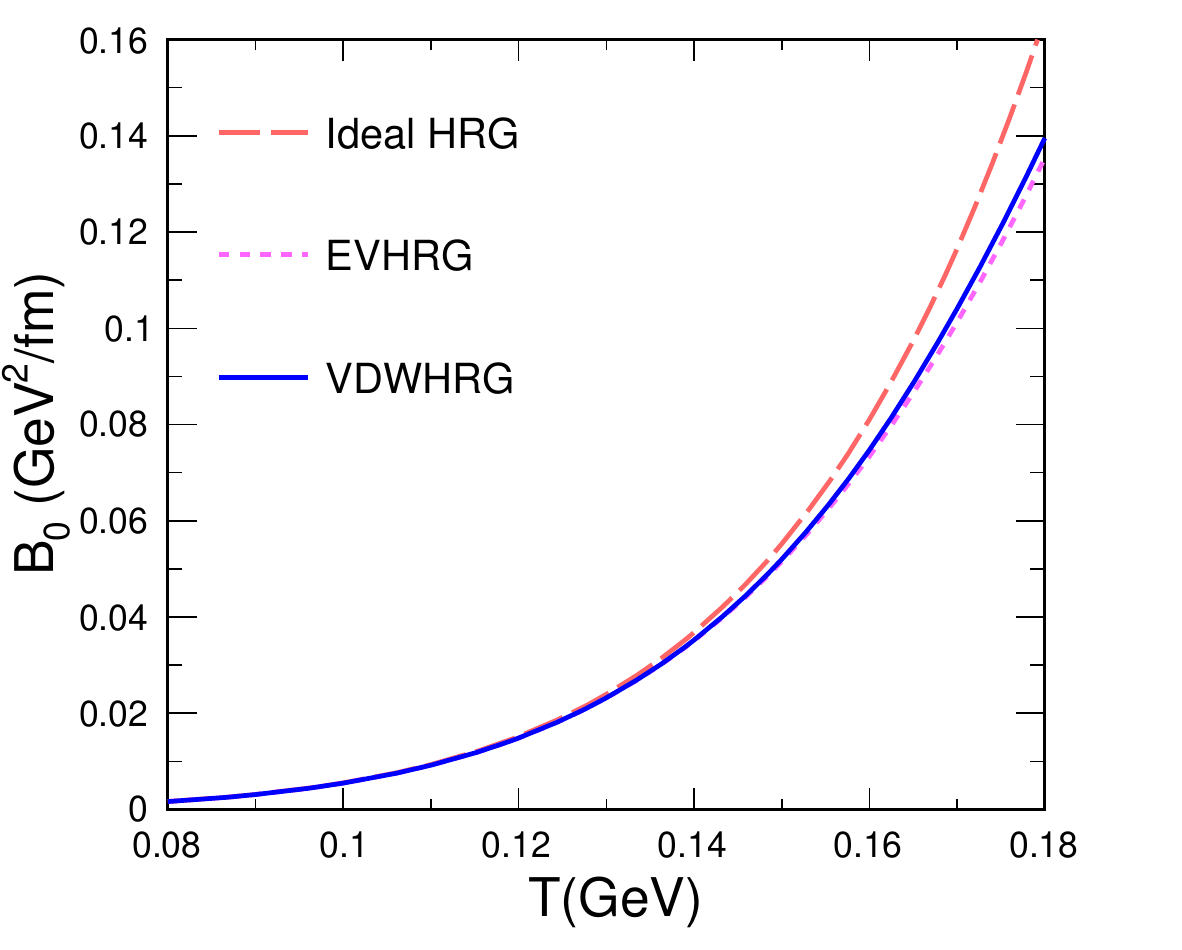}
    \includegraphics[width=0.45\linewidth]{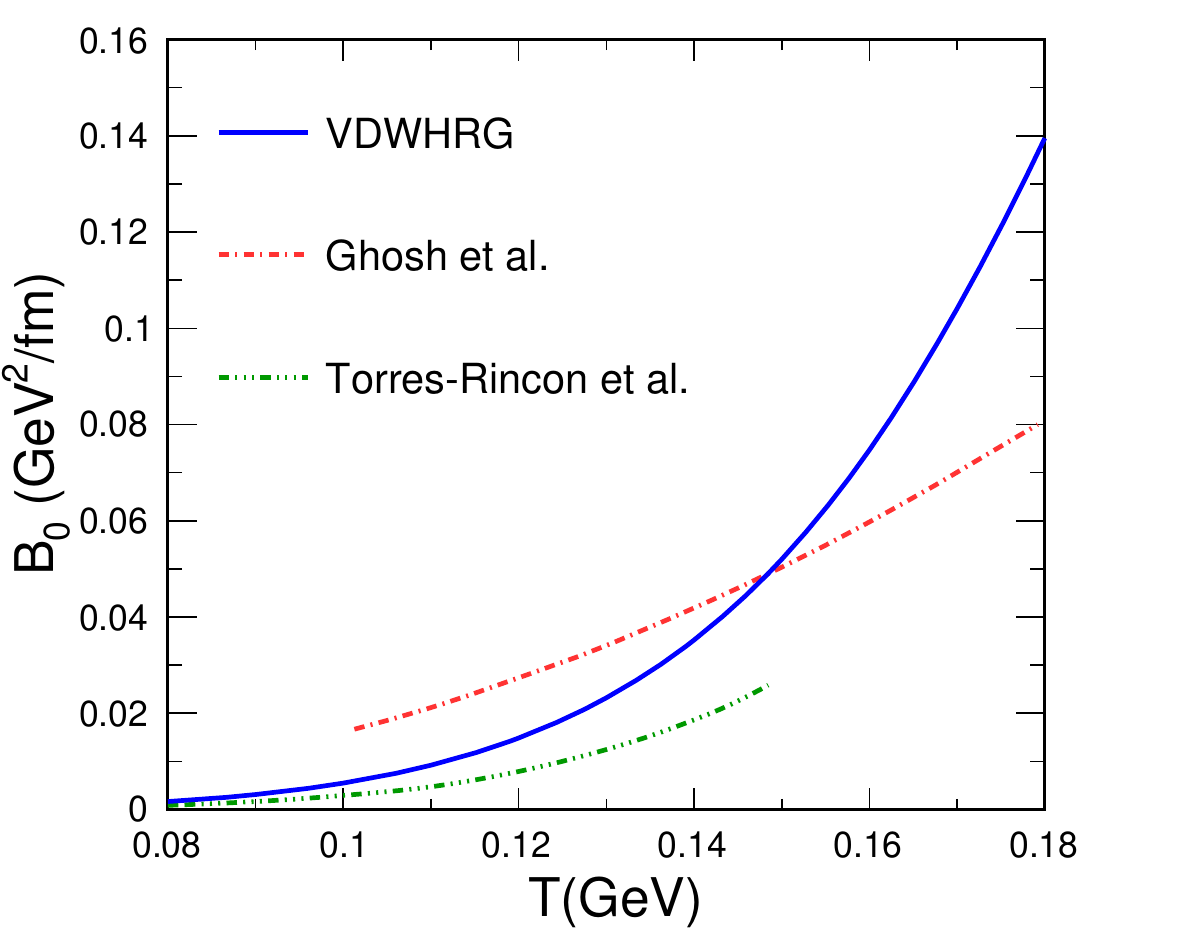}
    \caption{Transverse momentum diffusion coefficient of $D^{0}$ meson as a function of temperature at $\mu_{B}=0~\rm{GeV}$. A comparison among the Ideal HRG, EVHRG, and VDWHRG models (left). The red dot-dash line is the result of Ref.~\cite{Ghosh:2011bw} and the green dashed-dotted-dotted line is obtained from Ref.~\cite{Torres-Rincon:2021yga} (right); are compared with VDWHRG.}
    \label{fig3}
\end{figure*}

\begin{figure*}
    \centering
    \includegraphics[width=0.45\linewidth]{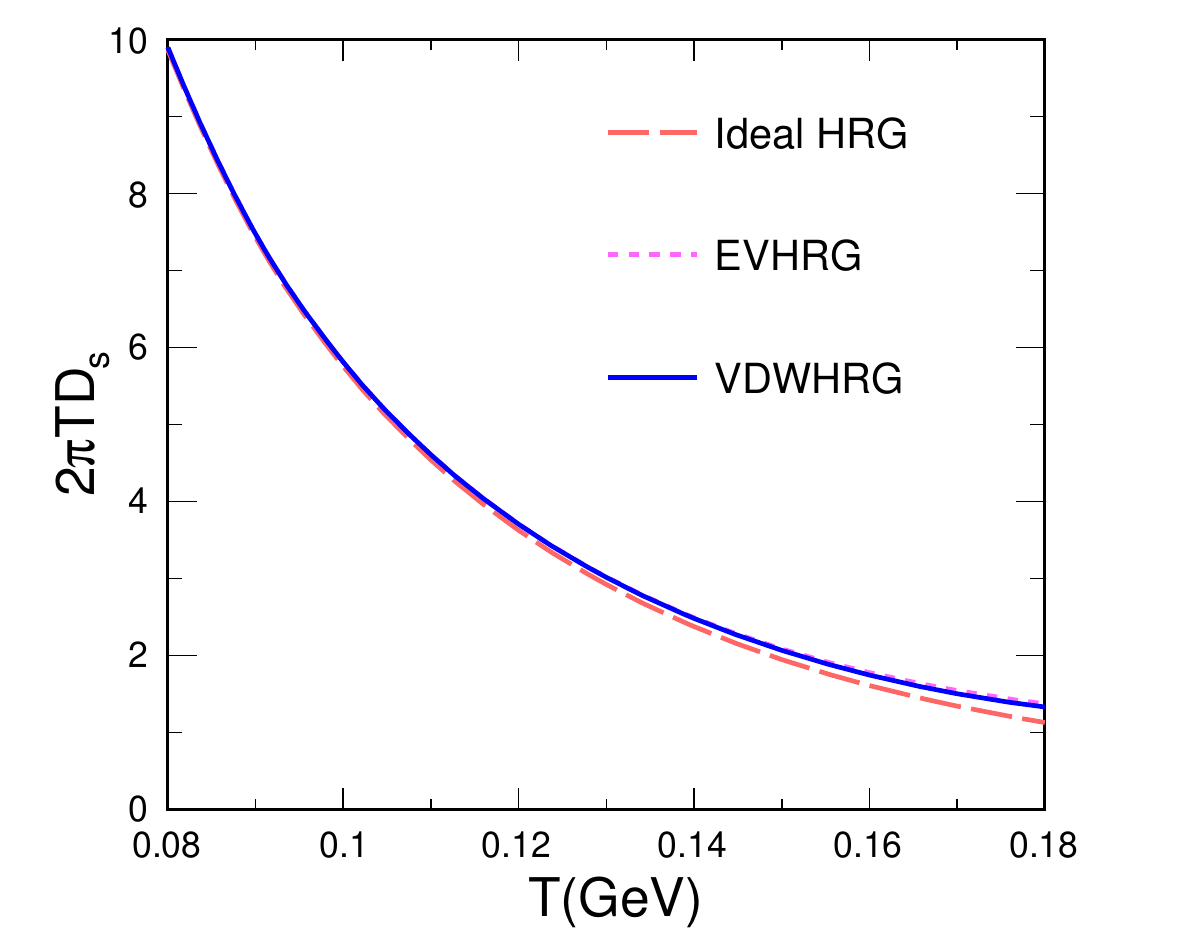}
    \includegraphics[width=0.45\linewidth]{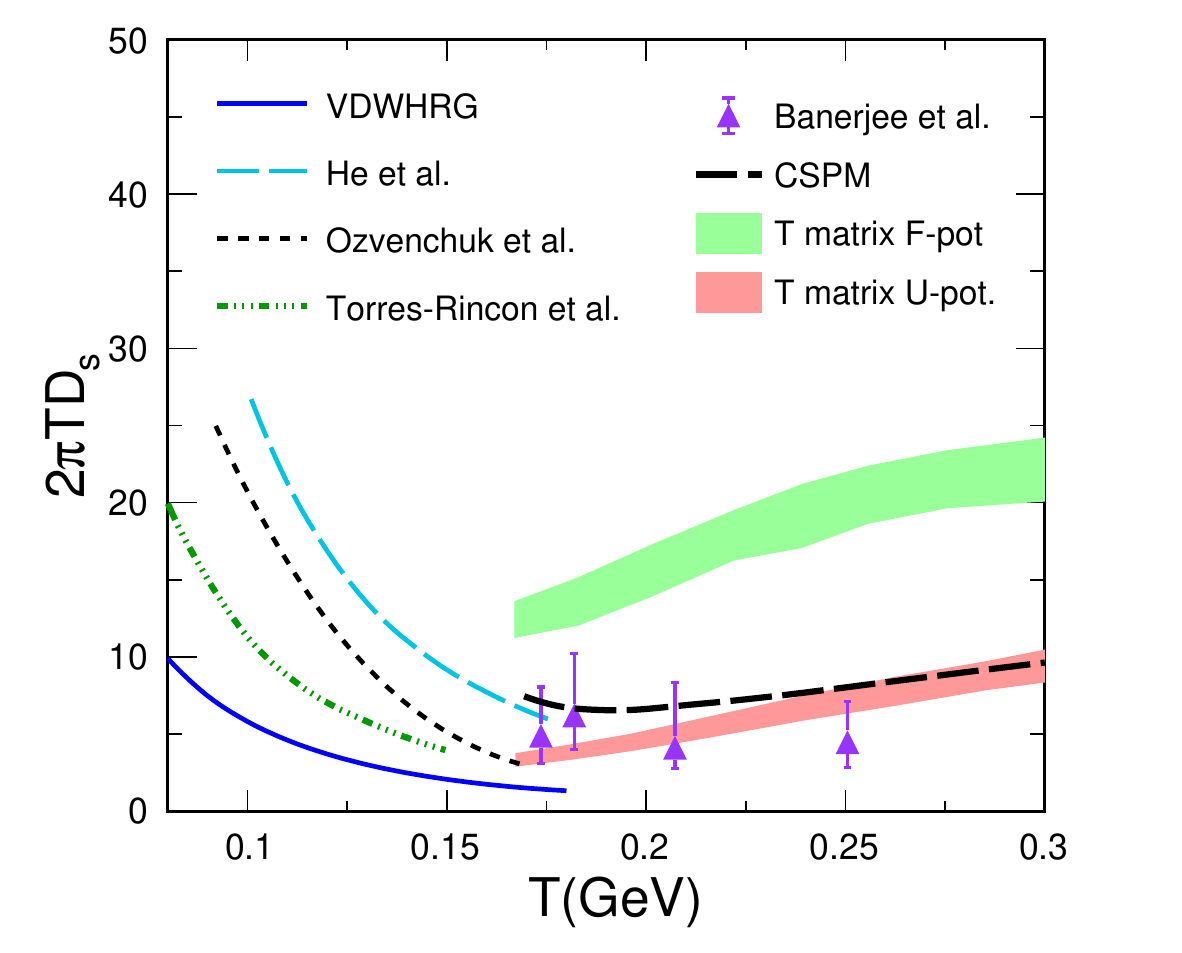}
    \caption{Spatial diffusion coefficient of $D^{0}$ meson as a function of temperature at $\mu_{B}=0~\rm{GeV}$. A comparison among the Ideal HRG, EVHRG, and VDWHRG models (left). A comparison of our findings with other phenomenological models (right). The violet markers are from lQCD calculations~\cite{Banerjee:2011ra}, the black dashed line is from Ref.~\cite{Goswami:2022szb}, and the green and red bands are obtained from T-matrix calculation~\cite{Riek:2010fk}. On the hadronic phase, the black dotted line is taken from Ref.~\cite{Ozvenchuk:2014rpa}, the dashed-dotted-dotted line is from Ref.~\cite{Torres-Rincon:2021yga} and the cyan dashed line is taken from~\cite{He:2011yi}.}
    \label{fig4}
\end{figure*}

\begin{figure*}
    \centering
    \includegraphics[width=0.32\linewidth]{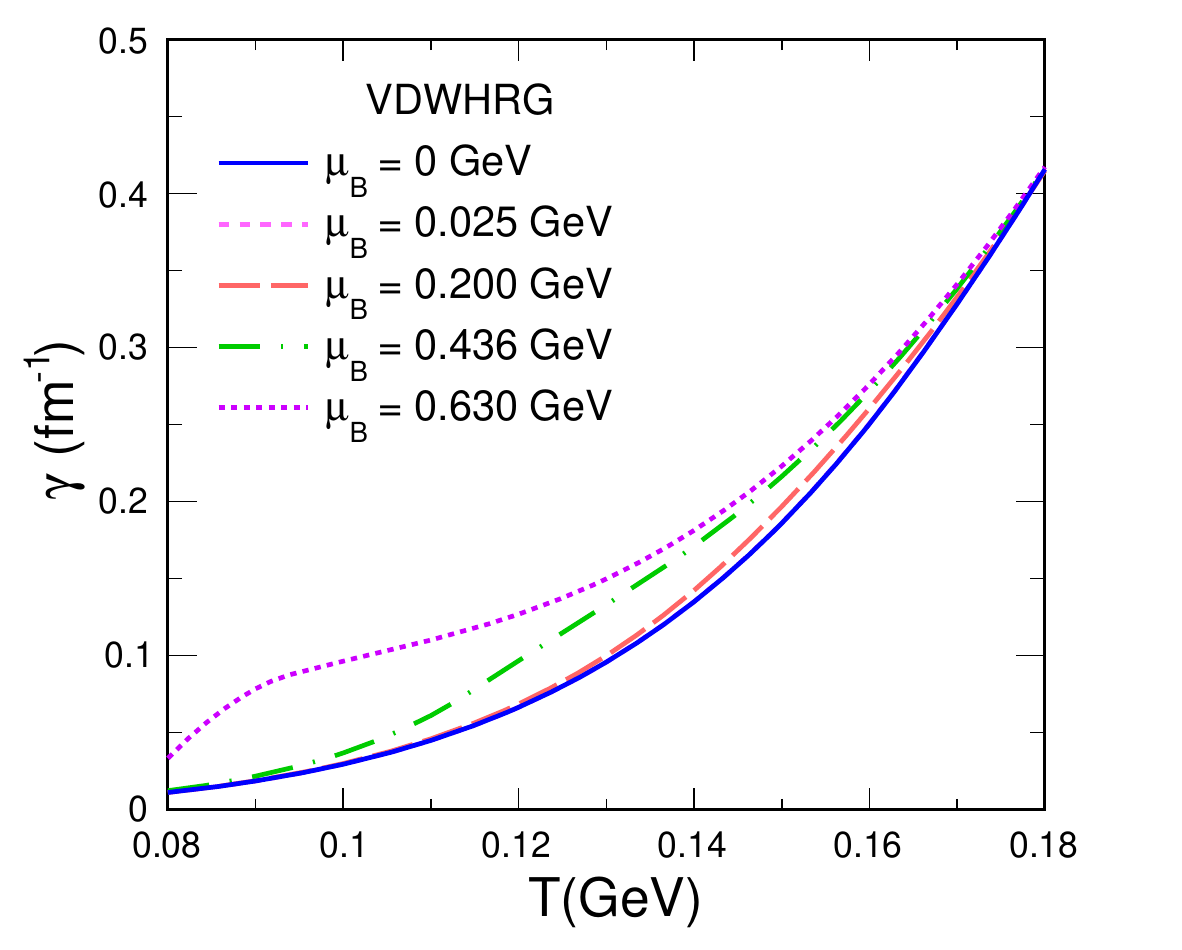}
    \includegraphics[width=0.32\linewidth]{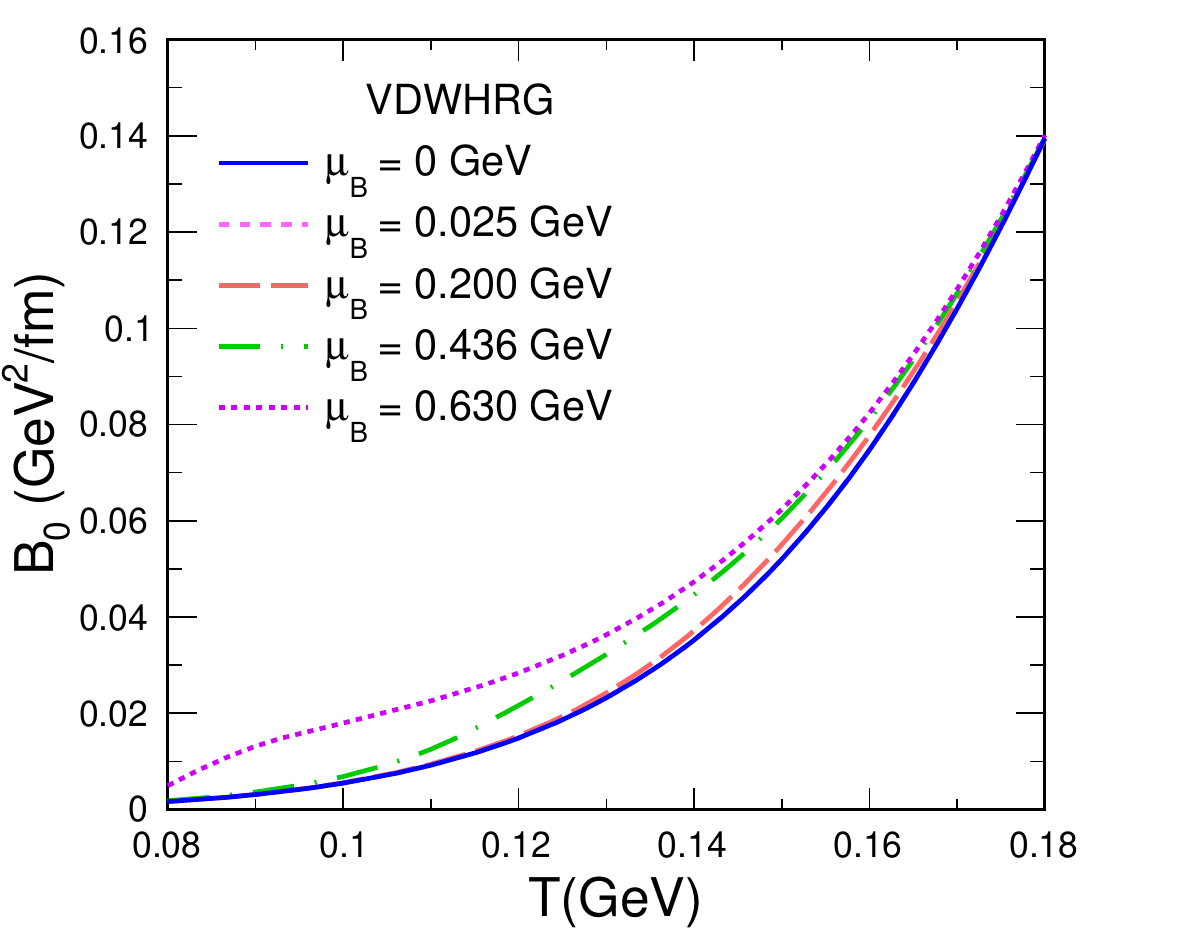}
    \includegraphics[width=0.32\linewidth]{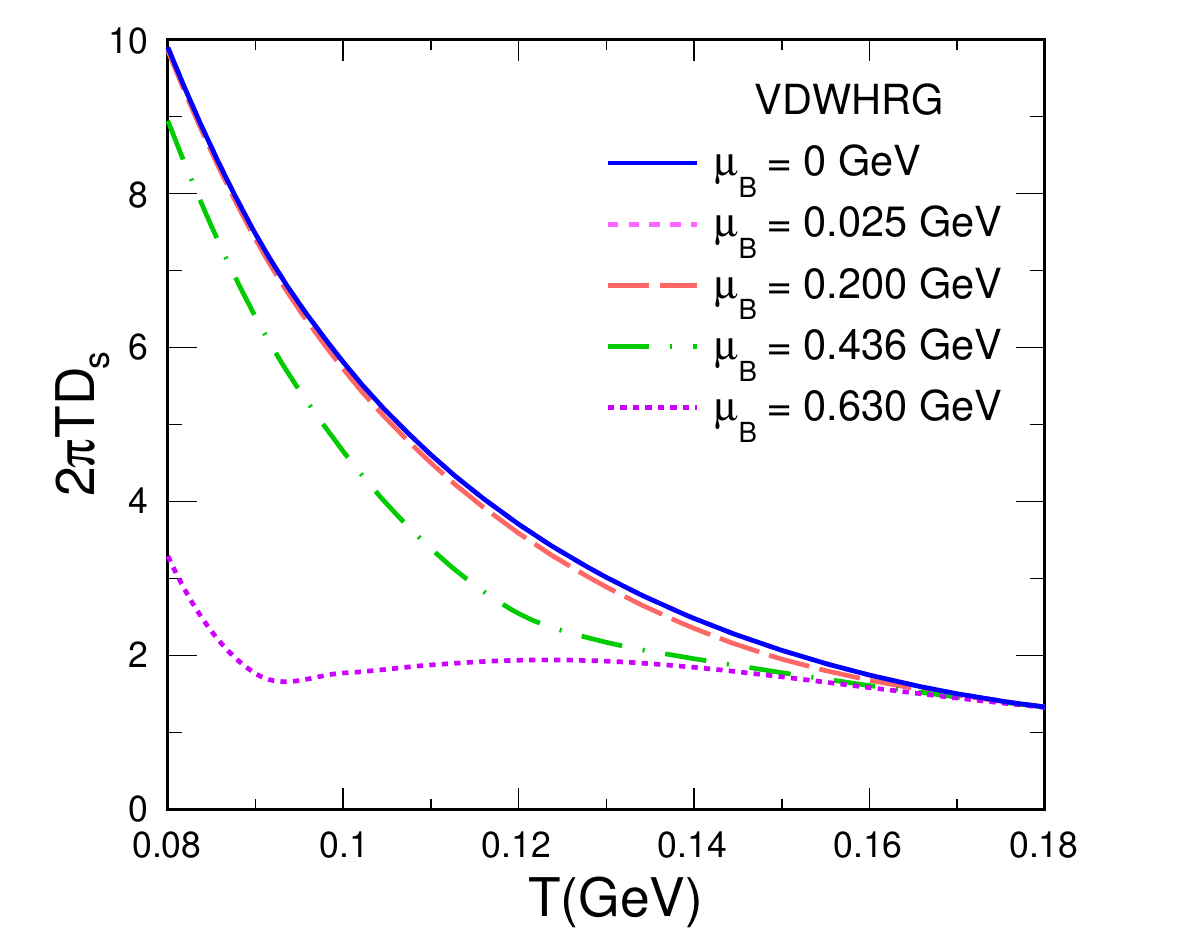}
    \caption{Drag coefficient, transverse momentum diffusion coefficient, and spatial diffusion coefficient as a function of temperature for finite $\mu_{B}$ values.}
    \label{fig5}
\end{figure*}

In fig.~(\ref{fig1}), we estimate the $D^{0}$ meson drag coefficient, $\gamma$, in a thermal bath with different mass cutoffs by using the Eq.\ref{eqtau}, where the number density is estimated using the van der Waals formalism. For only pion gas, the $D^{0}$ meson drag coefficient increases as a function of temperature. The trend remains the same for a gas of pions, kaons, and protons, but the magnitude of $\gamma$ increases. This is because the $D^{0}$ meson will undergo significantly more interactions, in a ($\pi+K+p$) gas as compared to a pion gas, resulting in an increase in $\gamma$. We also use a mass cutoff of 1 GeV for the hadrons in the medium, for that case the $D^{0}$ meson drag increases in magnitude as compared to the ($\pi+K+p$) gas. Finally, for a 1.2 GeV mass cutoff, it can be seen that the drag coefficient does not change much as compared to the previous 1.0 GeV mass cutoff case. This can be attributed to the negligible contribution of heavier hadrons due to their reduced number density in the hadron gas. For further calculations in this section, we use the mass cut-off of 1.2 GeV for the hadron gas.

In the left panel of Fig. \ref{fig2}, we study the variation of drag coefficient with temperature for various HRG models. We compare the values obtained from ideal HRG, EVHRG, and VDWHRG models. The ideal HRG model, where the number density of the hadronic medium is the highest, gives higher drag coefficient values. On the other hand, for the EVHRG model, due to the inclusion of the hardcore radius, the number density is suppressed. This causes the drag coefficient to be lower. However, the drag coefficient estimated using the VDWHRG model is slightly higher than the EVHRG model due to the attractive interaction, compensating for some repulsive effects. On the right panel of Fig.\ref{fig2}, we compare our VDWHRG estimation of the drag coefficient with different phenomenological models. Ghosh et al. \cite{Ghosh:2011bw} use an effective field theory to study the interaction between open charm mesons in a hot hadronic medium comprising of pions, nucleons, kaons, and eta mesons. Using the Kadanoff-Baym approach, Torres-Rincon et al. \cite{Torres-Rincon:2021yga} derived the off-shell Fokker-Planck equation that encodes the heavy-flavor transport coefficients. The drag and diffusion coefficients of $D^{0}$ mesons in a hadronic matter as a function of the momentum of $D^{0}$ mesons and the temperature of the medium at zero chemical potential have been computed by Ozvenchuk et al. \cite{Ozvenchuk:2014rpa}. In Ref.~\cite{He:2011yi}, authors estimate the drag coefficient using empirical elastic scattering amplitudes. It is observed that our estimation of the drag coefficient seems to agree well with the other models.

We study the variation of the transverse momentum diffusion coefficient as a function of temperature in Fig.~\ref{fig3}. The transverse momentum diffusion coefficient accounts for the broadening of the momentum spectra. A correlation has been observed between temperature, $T$, and the transverse momentum diffusion coefficient, $B_{0}$, showing that a rise in temperature increases the momentum broadening. From the left panel of Fig.~\ref{fig3}, we observe maximum momentum broadening of $D^{0}$ spectra in the thermal bath of an ideal hadron gas, while reduced number density reduces the momentum diffusion coefficient. Comparing our VDWHRG estimation with other phenomenological works in the right panel of Fig.~\ref{fig3}, we find that our estimation is consistent with the results reported by Ghosh et al. \cite{Ghosh:2011bw} and Torres-Rincon et al. \cite{Torres-Rincon:2021yga}.

In Fig.~\ref{fig4}, we compute the spatial diffusion coefficient and observe its variation with temperature. We observe a decreasing trend with an increase in temperature. The AdS/CFT calculation yields a lower bound of $2\pi T D_{s}$=1 near the critical temperature; as we approach $T_{c}$, we can see that the value of $2\pi TD_{s}$ tends towards a minimum. A slight difference in the estimation of $2\pi TD_{s}$ from different HRG models is due to the effect of interactions in the models. EVHRG estimates the maximum spatial diffusion coefficient; this can be understood as due to the repulsive interactions in the EVHRG model, the number density decreases significantly, allowing the $D^{0}$ meson to diffuse with relative ease. On the right panel, we compare our VDWHRG results with other studies estimating the spatial diffusion coefficient. In the hadronic medium, the results obtained by Torres-Rincon et al. \cite{Torres-Rincon:2021yga}, Ozvenchuk et al. \cite{Ozvenchuk:2014rpa} and He et al. \cite{He:2011yi} align well with our result and show a decrease with an increase in temperature. This is because as the temperature increases, the number density increases, and as a result, the interaction in the medium increases as well, which in turn, decreases the spatial diffusion coefficient. We can observe minima near the critical temperature owing to the emergence of a deconfined medium. In the partonic phase, our previous work \cite{Goswami:2022szb} and the result obtained from the T-matrix approach \cite{Riek:2010fk} show an increase in $D_{s}$ with increasing temperature. This is because, at higher temperatures, the partons will be asymptotically free, resulting in the weakening of strong interaction, which causes $2\pi TD_{s}$ to increase at higher temperatures.

In Fig.~\ref{fig5}, we study the drag coefficient, transverse momentum diffusion coefficient, and spatial diffusion coefficient of $D^{0}$ meson in a van der Waals HRG model for finite baryonic chemical potential. The $\mu_{B}$ values taken correspond to various colliders at different collision energies. $\mu_{B}$ = 0 GeV corresponds to the LHC, $\mu_{B}$ = 0.025 and 0.200 GeV correspond to RHIC at $\sqrt{s_{NN}}$ = 200 GeV and 19.6 GeV respectively. Similarly, $\mu_{B}$ = 0.436 and 0.630 GeV correspond to RHIC/FAIR at $\sqrt{s_{NN}}$ = 7.7 GeV and NICA at $\sqrt{s_{NN}}$ = 3 GeV respectively~\cite{Tawfik:2016sqd, Braun-Munzinger:2001hwo,Cleymans:2005xv, Khuntia:2018non}. We observe a trend of increasing values of $\gamma$ and $B_{0}$ with an increase in $\mu_{B}$. However, as the number density saturates at a higher temperature, the distinction between high and low $\mu_{B}$ becomes nonexistent. As $\mu_{B}$ increases, the spatial diffusion coefficient displays a similar pattern of decreasing value at lower and intermediate temperatures. However, at higher temperatures, $2 \pi T D_{s}$ approaches a consistent value regardless of the $\mu_{B}$ value. For $ \rm \mu_{B} = 0.630$ GeV, we observe a non-monotonic change in the value of $\gamma$, $B_{0}$, and $D_{s}$ at a lower temperature. It is more visible for the spatial diffusion coefficient, $2 \pi T D_{s}$. This might be due to the possible approach of the system towards the liquid-gas phase transition in the VDWHRG model. It is to be noted here that the critical point for the liquid-gas phase transition depends on the choice of VDW parameters. With the parameters obtained from ground state nuclear matter, $a = 329$ MeV fm$^3$ and $b = 3.42$ fm$^3$, the critical point lies at $T = 19.7$ MeV and $\mu_{B} = 908$ MeV \cite{Vovchenko:2015vxa}. In ref. \cite{Samanta:2017yhh}, the critical point is found at $T = 62.1$ MeV and $\mu_{B} = 708$ MeV corresponding to the parameters $a = 1250$ MeV fm$^3$ and $b = 5.75$ fm$^3$ (for $r = 0.7$ fm). In this work, the VDW parameters used are, $a = 926$ MeV fm$^3$, $r_{B(\bar B)} = 0.62$ fm and $r_{M} = 0.2$ fm, which are taken from the ref. \cite{Sarkar:2018mbk}, where the critical point is found to be around $T\simeq 65$ MeV and $\mu_{B}\simeq 715$ MeV. Therefore, we expect that if we go lower in temperature towards $65$ MeV and high $\mu_{B}$ upto $715$ MeV, the curve for $2 \pi T D_{s}$ may show the discontinuity due to the beginning of the liquid-gas phase transition.

\section{Charm abundances in the hadronic medium}
\label{fluc}
The critical temperature for the transition from hadronic to partonic degrees of freedom at zero baryochemical potential is estimated to be around 155 MeV from the lQCD calculations \cite{HotQCD:2014kol}. The light quark bound states dissolve at or around this temperature, demonstrating the strong connection between the chiral crossover and deconfinement of light quark degrees of freedom. This results in an abrupt change in the bulk thermodynamic observables, such as the speed of sound, which gives a minimum around $T_{\rm c}$ \cite{HotQCD:2014kol}. This change is even more evident in the behavior of fluctuations of conserved charges, such as baryon number, electric charge, or strangeness fluctuations. The quick shift in the degrees of freedom carrying the necessary conserved charges is directly reflected in the ratios of the various moments (cumulants) of net-charge fluctuations and their correlations in the transition zone. The overall number of hadronic degrees of freedom or the precise hadronic mass spectrum also affects bulk thermodynamics. To illustrate, the high rise of the trace anomaly, which was discovered in lattice QCD computations, may indicate contributions from hadron resonances that have not yet been seen \cite{Sharma:2013hsa}. In addition, some recent works have shown that the chemical freeze-out temperature for the light hadrons is not the same as that of the strange hadrons \cite{Bellwied:2013cta, Flor:2020fdw}. The single strange, doubly strange, triply strange hadrons all freeze out at different temperatures, thus emphasizing the case for a differential chemical freeze-out scenario~\cite{Flor:2021olm, Bellwied:2017uat}. In the same line, one can assume that such a condition may be observed in the charm sector as well. Thus, a thorough investigation is necessary for the charm sector.

\begin{figure}
    \includegraphics[width=1\linewidth]{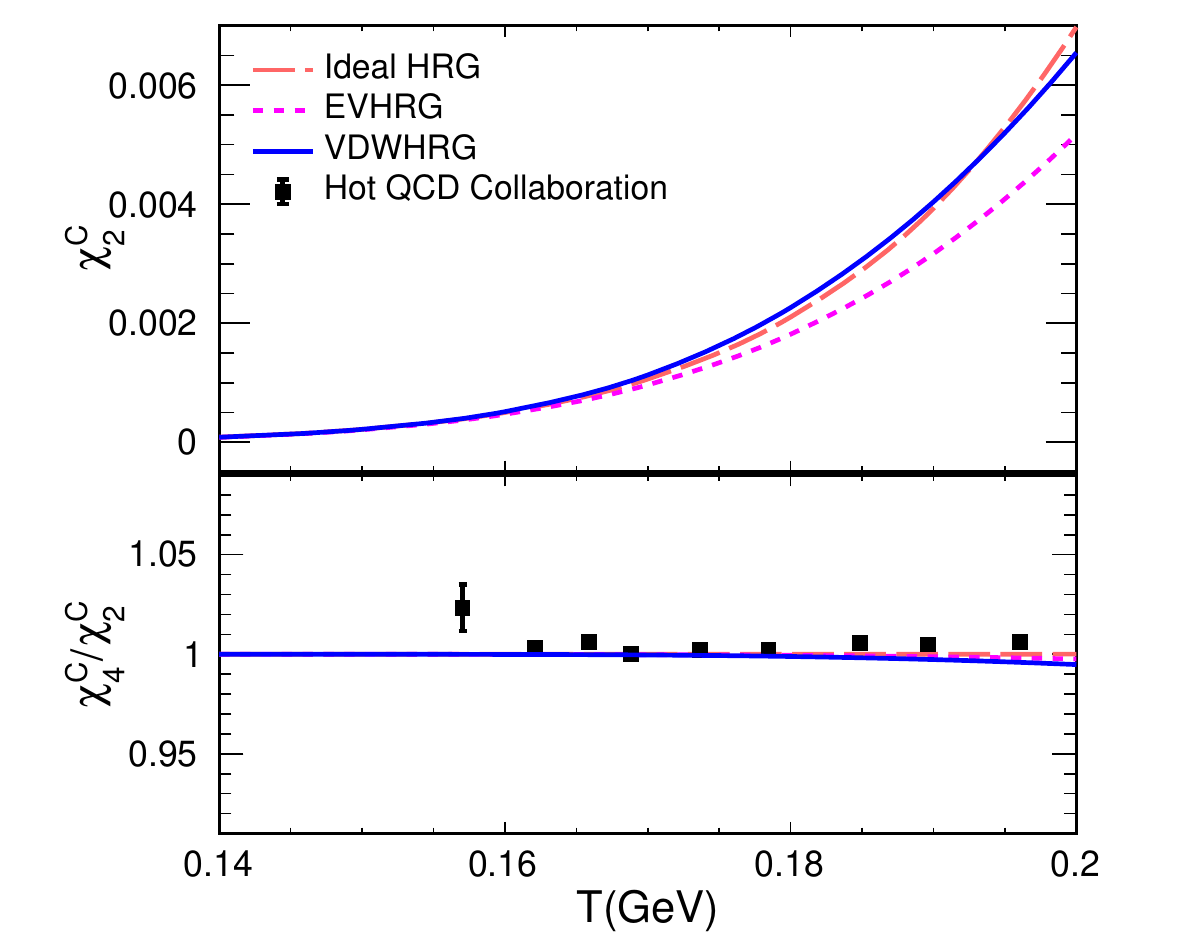}
    \caption{Variation of $\chi_{2}^{C}$ with temperature (top) and the ratio $\chi_{4}^{C}$/$\chi_{2}^{C}$ (bottom) compared with Hot QCD collaboration \cite{Bazavov:2014yba}.}
    \label{fig6}
\end{figure}

\begin{figure*}
    \includegraphics[width=0.45\linewidth]{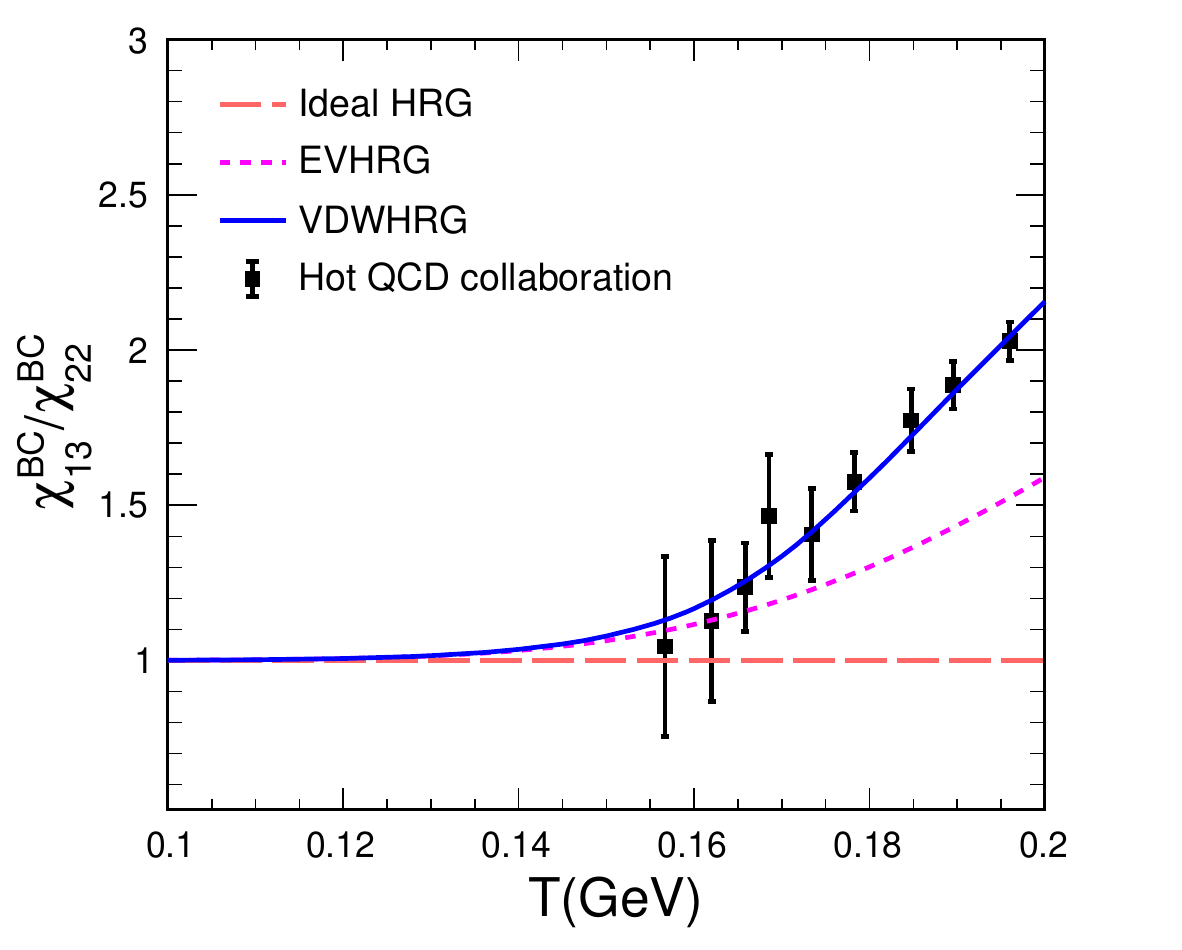}
    \includegraphics[width=0.45\linewidth]{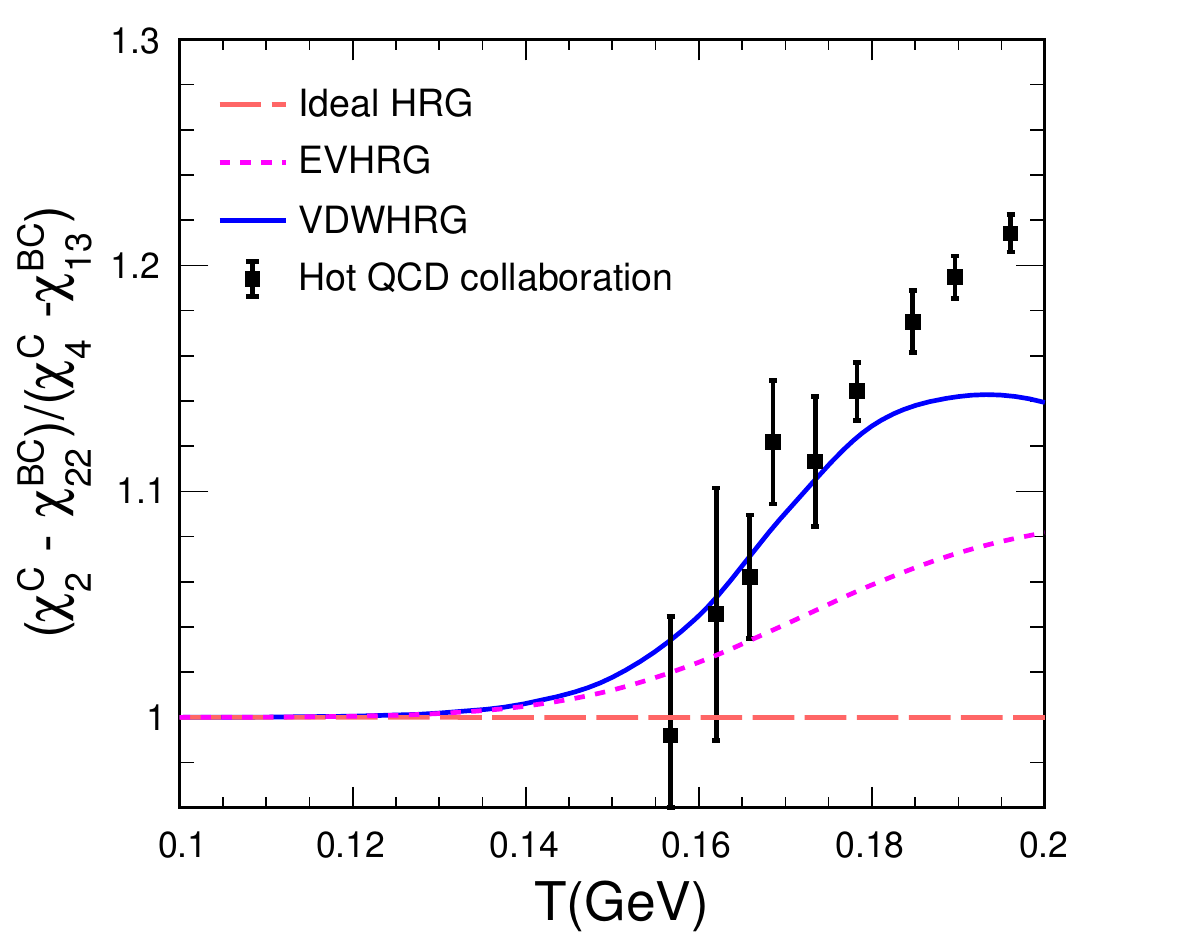}
    \caption{Second-order and fourth-order fluctuation as a function of temperature computed using IHRG, EVHRG, and VDWHRG model. On the right panel, the ratio of fourth-order net baryon and charm correlations. On the left panel, we compute the contributions coming from the open charm mesons and a comparison with results from~\cite{Bazavov:2014yba}.}
    \label{fig7}
\end{figure*}

\begin{figure*}
    \centering
    \includegraphics[width=0.32\linewidth]{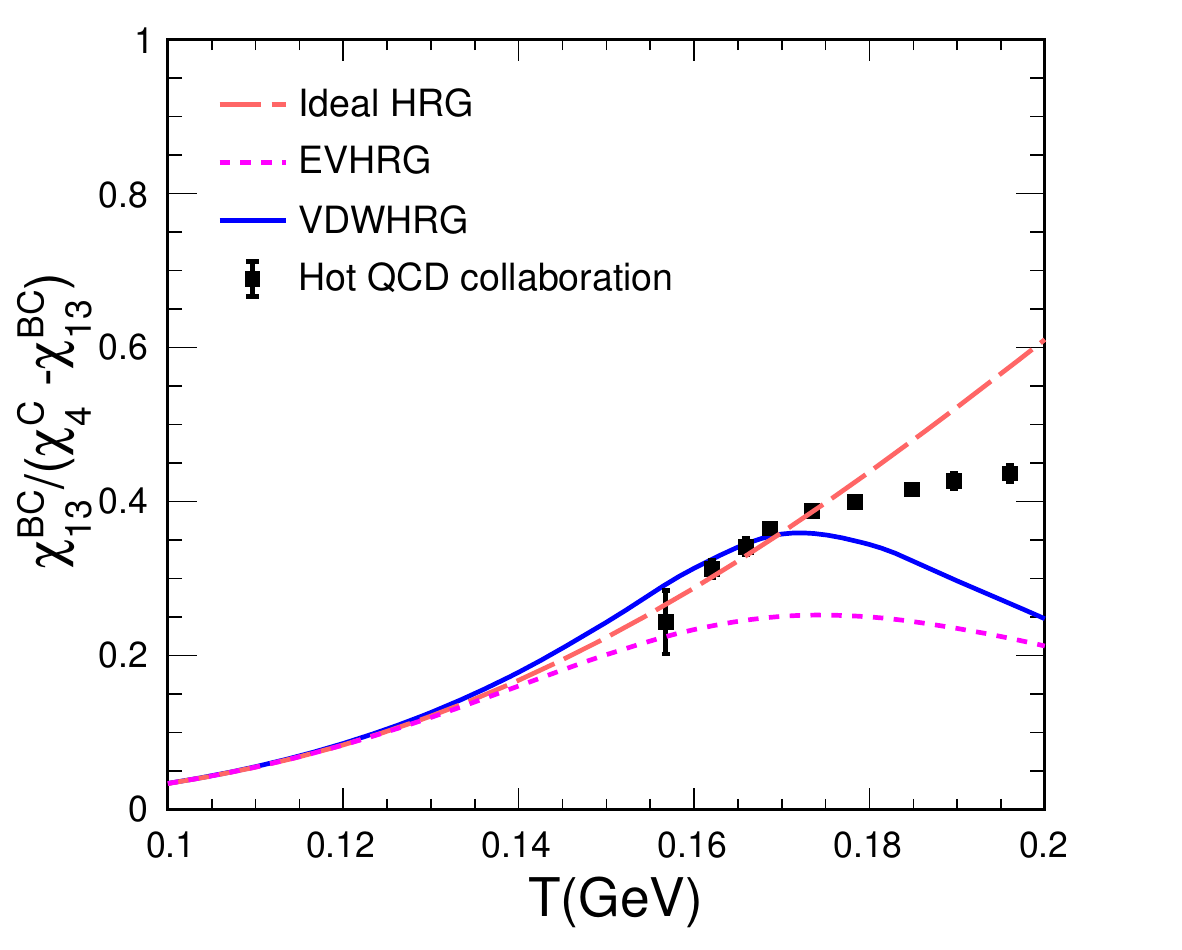}
    \includegraphics[width=0.32\linewidth]{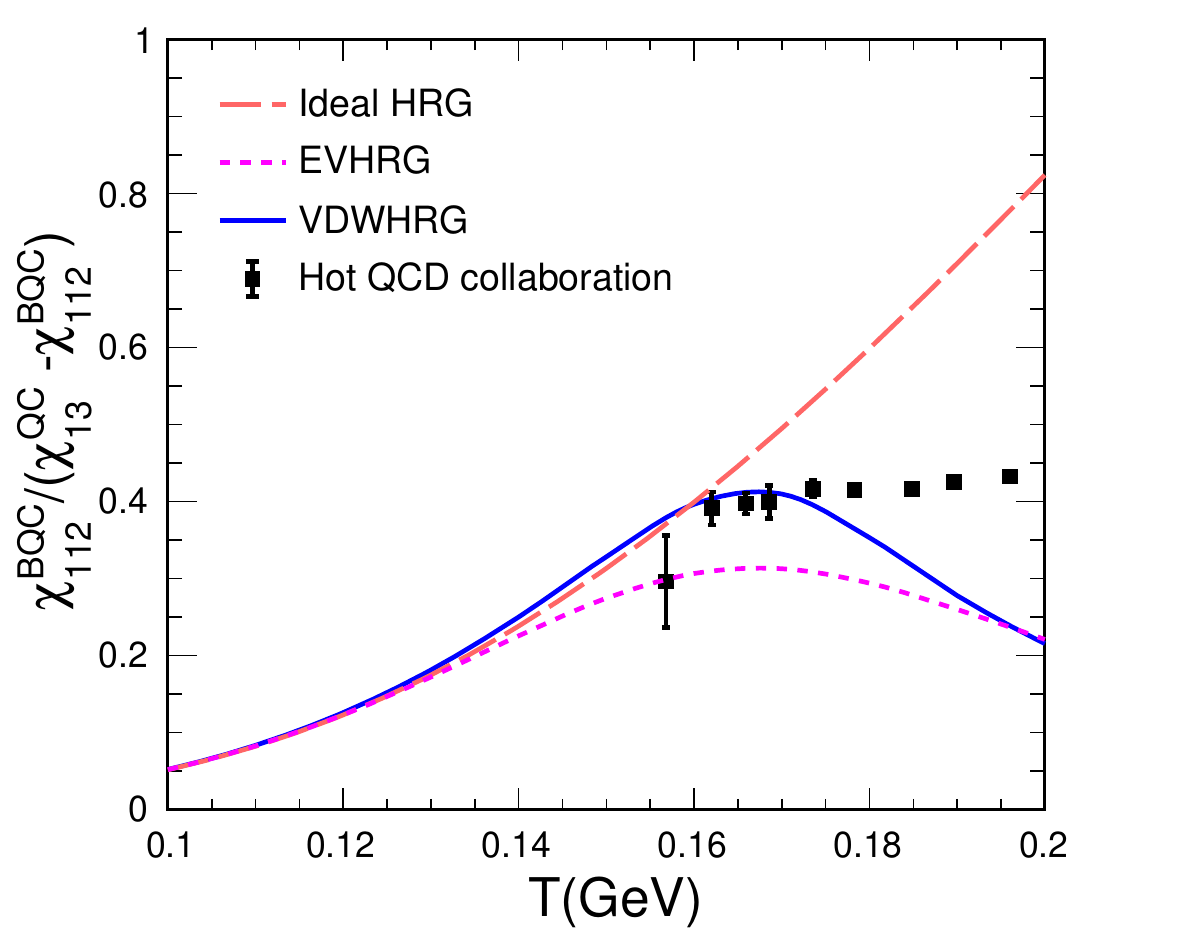}
    \includegraphics[width=0.32\linewidth]{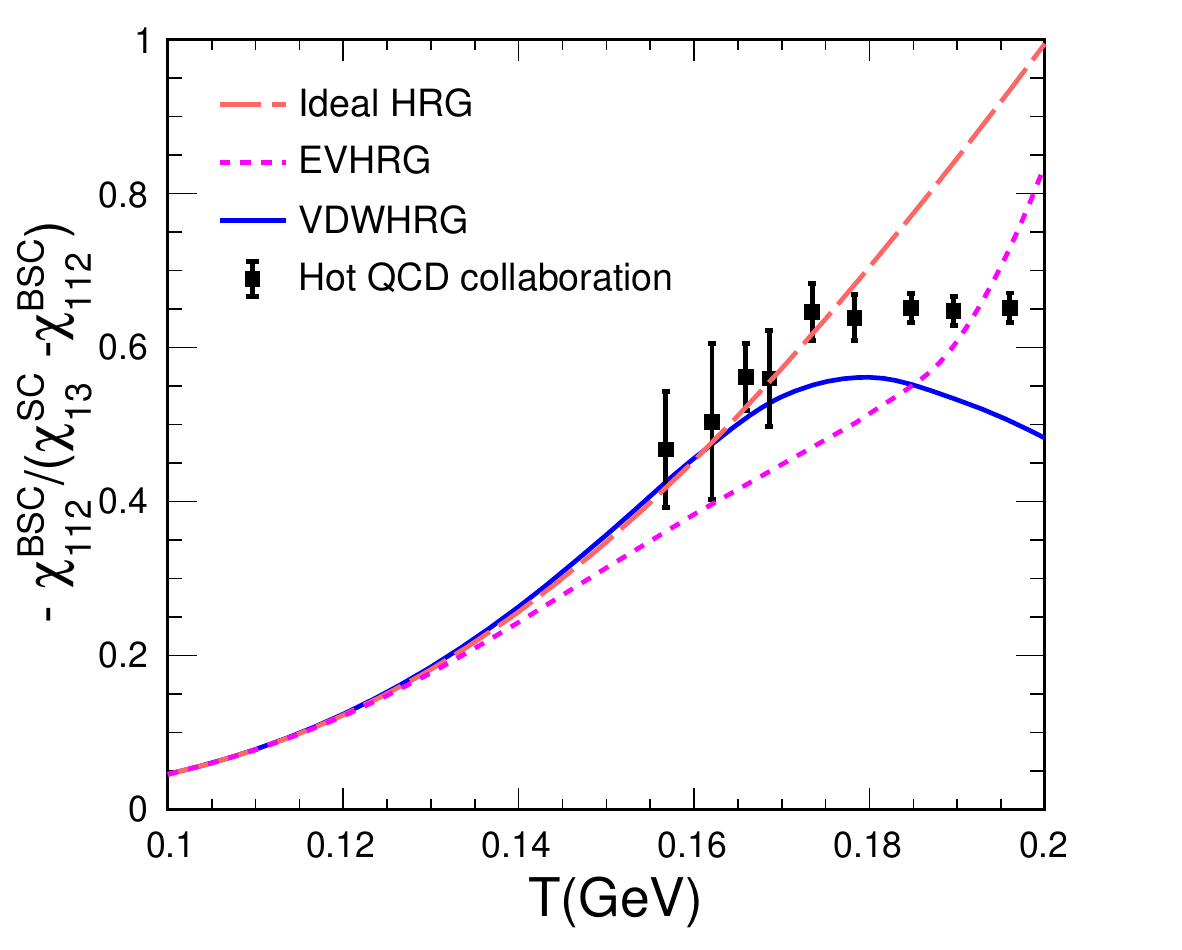}
    \caption{Fourth-order net charm fluctuation as a function of temperature computed using IHRG, EVHRG, and VDWHRG model. The ratio of fluctuations receiving contributions from all charmed hadrons (left), charged charmed hadrons (middle), and strange charmed hadrons (right) compared with data from Hot QCD collaboration \cite{Bazavov:2014yba}.}
    \label{fig8}
\end{figure*}

Although it appears to be established that charmonium states, or bound states with hidden charm, continue to exist in the QGP at temperatures much higher than $T_{\rm c}$~\cite{Rapp:2008tf}, this may not be true for the heavy-light mesons or baryons, such as open charm mesons ($D^{0}, D^{+}, D^{-}, D_{s}$) or charmed baryons ($\Lambda_{c}, \Xi_{c}, \Omega_{c}$)~\cite{Bazavov:2014yba}. To answer this query of melting charmed hadrons, one needs to compute net-charm fluctuations, cumulants, and correlations between their moments and moments of net baryon number, electric charge, or strangeness fluctuations. We can compute the susceptibilities of the conserved charges by the formula given by
\begin{equation}
    \chi^{\rm BSQC}_{\rm ijkl} = \frac{\partial^{\rm i+j+k+l}(P/T^{4})}{\partial(\mu_{B}/T)^{\rm i}(\mu_{S}/T)^{\rm j}(\mu_{Q}/T)^{\rm k}(\mu_{C}/T)^{\rm l}}.
\end{equation}
Mathematically, an $n^{th}$ order cumulant is an $n^{th}$ order derivative of pressure with respect to corresponding chemical potential. Such an $n^{th}$ order derivative of pressure with respect to $\mu_{C}$, gives us a quantity with a coefficient $C^{n}$. Thus, one can take suitable ratios of such cumulants to find the appropriate ratios of desired quantum numbers. For this section of the calculation, we use the particle list from the Particle data group (PDG) \cite{ParticleDataGroup:2016lqr}. In addition, we have included the undiscovered charmed states predicted by the quark model \cite{Ebert:2009ua, Ebert:2011kk}, without which one underestimates the lQCD data \cite{Bazavov:2014yba}. In the upper panel of Fig.~\ref{fig6}, we study the variation of second-order, ($\chi_{2}^{C}$), cumulant with temperature. We observe an increase in their values with an increase in temperature. This is primarily because, numerically, second-order fluctuation is the second derivative of pressure with respect to the corresponding chemical potential. As the pressure increases, the second-order susceptibility increases as well. A slight deviation can be observed between the IHRG, VDWHRG, and EVHRG estimations. In the lower panel of Fig. \ref{fig6}, we plot the ratio of fourth-order net charm fluctuation to second-order net charm fluctuation. This ratio gives the kurtosis of the net-charm distribution. From the lQCD calculations, the ratio is estimated to be unity within errors, which means that the distribution is normal. Our results are in line with the findings from the lQCD calculation. One can see a slight deviation of the VDWHRG results from the HRG results at high temperatures.

In the left panel of Fig.~\ref{fig7}, we plot the ratio of fourth-order cumulants, $\rm \chi_{13}^{BC}$ and $\rm \chi_{22}^{BC}$. Mathematically, it can be understood as the ratio of charm number to baryon number. In the hadronic sector, we have $\rm |B|=|C|=1$, so the ratio should ideally be unity in an uncorrelated hadronic medium. In the partonic phase, we have $\rm|C|=1$ and $\rm |B|=1/3$, thus the ratio rises to 3. For this study, as mentioned in \cite{Bazavov:2014yba}, we have only considered hadrons with $\rm |C|=1$. This is due to the reason that hadrons with $\rm |C|=2$ and $\rm |C|=3$ are much heavier and their contribution is negligible. Our results depict that an ideal HRG model fails to explain the trend, while estimations from the VDWHRG model are consistent with lQCD data. The VDWHRG model can explain the lQCD data up to 200 MeV. The van der Waals interactions between the hadrons with increasing temperature mimic the behavior of a deconfined medium up to a certain temperature. An analogy can be drawn by understanding the PNJL model, where the quarks gain masses below the critical temperature, and the model can explain the hadronic sector even though no hadrons are present in the model \cite{Ratti:2005jh}. On the right panel of Fig.~\ref{fig7}, we plot the ratio of cumulants that receive contributions from the open charmed mesons. We calculate the contribution due to the charmed meson fluctuation from any second-order or fourth-order charmed fluctuation by subtracting the contribution of the charmed baryons. In this plot, one observes that the HRG model again fails to explain the rise in the ratio of the cumulant with temperature. However, findings from the VDWHRG model are consistent with lQCD up to around 180 MeV. 

Finally, in Fig.~\ref{fig8}, we compute fourth-order cumulants of net charm fluctuations and their correlation with conserved charges like net baryon number, electric charge, and strangeness. We take ratios of appropriate cumulants sensitive to the melting of charmed hadrons and study their dependence on temperature. On the left panel, we plot the ratio of net charmed baryon fluctuation to net charmed meson fluctuation. The middle panel shows the fluctuation ratio of all charged-charmed baryons to all charged-charmed mesons. On the right panel, we show the variation in the ratio of all strange-charmed baryons to all strange-charmed mesons. We observe a similar trend in all the plots: the ideal HRG model increases almost monotonically with temperature, deviating from the lQCD results after temperatures 160-170 MeV. However, results from the VDWHRG model reach a peak of around 160-170 MeV and then decrease towards higher temperatures. The disagreement between the lQCD and VDWHRG at high temperatures may be due to the reason that in our study, we have excluded the meson-meson attraction, meson-anti(baryon) interactions, and baryon-anti baryon interactions. However, one can clearly see that the VDWHRG model explains the lQCD data very well compared to the IHRG and EVHRG models.

\section{Summary and Discussion}
\label{sum}

In this work, we present a phenomenological estimation of the drag and diffusion coefficients of the $D^{0}$ meson. We also study the effect of baryon chemical potential and the interactions in the system on the $D^{0}$ meson diffusion. We estimate the drag force experienced by $D^{0}$ meson in a pion gas, $(\pi+K+p)$ gas, and in a hadron gas with mass cutoffs 1 and 1.2 GeV at $\mu_{B}=0~\rm{GeV}$. $D^{0}$ meson interact substantially more in a denser medium which leads to a higher drag force. Further, we study the temperature dependence of the transverse momentum diffusion coefficient, $B_{0}$, which accounts for the broadening of the final state particle momentum spectra. One can observe a linear increase of $B_{0}$ with temperature. Such an increase can be attributed to the fact that number density increases with temperature, which enables $D^{0}$ mesons to diffuse more in momentum space. Finally, we study the spatial diffusion of $D^{0}$ meson, scaled by a factor of $2\pi T$. Our estimation of $2\pi T D_{s}$ approaches the minima around the critical temperature. Our results are compatible with various phenomenological models. We also study the effect of finite baryon chemical potential on the drag and diffusion coefficients. Due to the van der Waals interaction, a non-monotonic behavior can be seen at low temperatures and high-$\mu_{B}$ regime.

Moreover, we estimate the charm number fluctuations within the van der Waals hadron resonance gas model. By taking an appropriate ratio of cumulants we study the melting of open-charmed hadrons. The ideal HRG model fails to explain the lQCD data, whereas by introducing VDW interactions, we observe that our results show a good agreement with the lQCD results up to $T \simeq $ 180 MeV. This study can help us to understand the melting of charmed hadrons in a hot and dense medium formed in an ultra-relativistic collision.

The study of heavy-flavour hadron dynamics provides us with unique opportunities to understand the hot and dense matter produced in heavy-ion collisions at ultra-relativistic energies. The $D^{0}$ meson, which is the lightest neutral charmed hadron, can give us information about the medium through its study of the drag and diffusion coefficients. This is encoded within the elliptic flow ($v_{2}$) and the nuclear suppression factor ($R_{AA}$) of $D^{0}$ meson, which can be measured in experiments. On the other hand, one can, in principle, study the net charm number fluctuation by taking net $D^{+}$ and $D^{-}$ meson fluctuations as a proxy. This is because, for net charm cumulants calculation, one needs to take two particle species (particle and antiparticle, $\Delta N_{c} = N_{c} - \bar N_{c}$) into consideration. By default, $D^{0}$ and $\bar D_{ 0}$ should have been the ideal choice, as they are the lightest charmed hadrons. However, as observed in the LHCb experiment with a significance of 8.2 standard deviations, the $D^{0}$ and $\bar D^{0}$ suffer from oscillations \cite{LHCb:2016zmn}. Hence, it is not an ideal probe for net charm cumulants estimation. However, looking at net $D^{+}$ and $D^{-}$ meson fluctuations, one can use them as probes to study charm number fluctuations at both ALICE and STAR experiments. In view of ALICE run-3 and a high luminosity collision environment, the charm sector will be of high importance. Results such as $D^{0}$ meson $v_{2}$ and $R_{AA}$ will be more accurate with smaller uncertainities. This means the theoretical and phenomenological models must be fine-tuned to explain the data. In addition, with higher statistics, it would be interesting to see the charm number cumulants, which would shed light on the melting of charmed hadrons in a hot and dense medium.

\section{Acknowledgement}
K.G. acknowledges the financial support from the Prime Minister's Research Fellowship (PMRF), Government of India. K.K.P. acknowledges the doctoral fellowships from the University Grants Commission (UGC), Government of India. The authors acknowledge the valuable discussion with Ronald Scaria. The authors gratefully acknowledge the DAE-DST, Government of India funding under the mega-science project “Indian participation in the ALICE experiment at CERN” bearing Project No. SR/MF/PS-02/2021-IITI(E-37123).

\end{document}